\documentclass[letterpaper]{article} 
\usepackage{aaai25}  
\usepackage{times}  
\usepackage{helvet}  
\usepackage{courier}  
\usepackage[hyphens]{url}  
\usepackage{graphicx} 
\usepackage{natbib}  
\usepackage{caption} 
\usepackage{algorithm}
\usepackage{algorithmic}

\usepackage{amsmath}
\usepackage{amssymb}
\usepackage{mathtools}
\usepackage{amsthm}

\usepackage{multirow}
\usepackage{listings}
\usepackage{xcolor}
\definecolor{codeblue}{rgb}{0.25,0.5,0.5}
\definecolor{codekw}{rgb}{0.85, 0.18, 0.50}
\usepackage[capitalize,noabbrev]{cleveref}

\usepackage{newfloat}
\usepackage{listings}
\DeclareCaptionStyle{ruled}{labelfont=normalfont,labelsep=colon,strut=off} 
\floatstyle{ruled}
\newfloat{listing}{tb}{lst}{}
\floatname{listing}{Listing}
\title{Improving Generalization for AI-Synthesized Voice Detection}

\author{
    Hainan Ren\thanks{Work done as a remote student under Prof. Hu's supervision},
    Li Lin\textsuperscript{\rm 1},
    Chun-Hao Liu\textsuperscript{\rm 2},
    Xin Wang\textsuperscript{\rm 3},
    Shu Hu\textsuperscript{\rm 1}\thanks{Corresponding author (hu968@purdue.edu)}
}
\affiliations{
    \textsuperscript{\rm 1} Purdue University\quad
    \textsuperscript{\rm 2} Amazon\quad
    \textsuperscript{\rm 3} University at Albany, SUNY\quad\\
    hnren666@gmail.com, lin1785@purdue.edu, chunhaol@amazon.com, xwang56@albany.edu, hu968@purdue.edu
}

\usepackage{bibentry}

\makeatletter
\DeclareRobustCommand\onedot{\futurelet\@let@token\@onedot}
\def\@onedot{\ifx\@let@token.\else.\null\fi}

\def\eg{\emph{e.g}\onedot} 
\def\ie{\emph{i.e}\onedot}

\makeatother

\newcommand{\baseline}{}
\newcommand{\tablestyle}[2]{
    \def\tabcolsep{#1}%
    \def\arraystretch{#2}%
    \centering\footnotesize
}

\def\sign{\texttt{sign}}

\usepackage{colortbl}

\begin{document}

\maketitle

\begin{abstract}
AI-synthesized voice technology has the potential to create realistic human voices for beneficial applications, but it can also be misused for malicious purposes. 
While existing AI-synthesized voice detection models excel in intra-domain evaluation, they face challenges in generalizing across different domains, potentially becoming obsolete as new voice generators emerge.
Current solutions use diverse data and advanced machine learning techniques (\eg, domain-invariant representation, self-supervised learning), but are limited by predefined vocoders and sensitivity to factors like background noise and speaker identity.
In this work, we introduce an innovative disentanglement framework aimed at extracting domain-agnostic artifact features related to vocoders. 
Utilizing these features, we enhance model learning in a flat loss landscape, enabling escape from suboptimal solutions and improving generalization.
Extensive experiments on benchmarks show our approach outperforms state-of-the-art methods, achieving up to 5.12\% improvement in the equal error rate metric in intra-domain and 7.59\% in cross-domain evaluations. 
\end{abstract}
\begin{links}
\link{Code}{https://github.com/Purdue-M2/AI-Synthesized-Voice-Generalization}
\end{links}
%

\section{Introduction}

\begin{figure}[ht]
\centering
\centerline{\includegraphics[width=1\columnwidth]{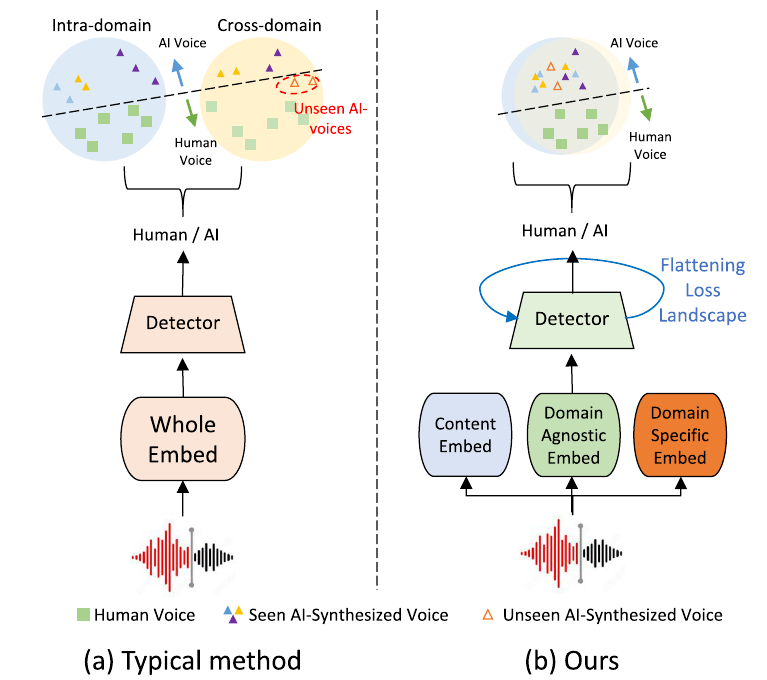}}
\caption{{Comparison of audio deepfake detection methods. 
    (a) The conventional method uses entire voice features and excels in distinguishing between human and AI-synthesized voices within familiar domains but struggles with voices generated from unseen vocoders, leading to inaccurate separation. 
    (b) Our method achieves intra-\&cross-domain detection by exposing domain-agnostic features for learning on a flattened loss landscape. 
    }
    }
\label{fig:intro-i}
\end{figure}

AI voice technology uses advanced models to generate natural human speech, mimicking tone and pronunciation to the extent that it is indistinguishable from real recordings.
This technology has a wide range of applications, including voice assistants, audiobooks, voiceovers for videos and advertisements, and in fields such as healthcare and accessibility, where it can assist individuals with speech impairments.
AI voice synthesis is rapidly evolving, using advanced variational auto-encoder (VAE) \cite{peng2020natts},  generative adversarial network (GAN) models \cite{kumar2019melgan, yamamoto2020parallel, lee2022fregan2}, and diffusion models \cite{chen2020wavegrad, kong2020diffwave, liu2023audioldm} to generate audio waveforms from mel-spectrograms that are nearly indistinguishable from human recordings.

However, the widespread use of AI-synthesized voice has raised concerns about potential misuse, such as creating fake voice recordings, or `Audio Deepfakes', for impersonation or fraud. For instance, in 2019, fraudsters used AI to mimic a CEO’s voice and stole over \$243,000 via a phone call \cite{Forbes}.
To address these concerns, research into detecting and mitigating AI-synthesized voice has become essential, with much of the progress \cite{barrington2023single,sun2023ai} driven by ASVspoof challenges and datasets \cite{asvspoof}. While these methods show promising results in intra-domain evaluations (\ie, training and testing data come from the same vocoder), they face significant performance drops in cross-domain testing (\ie, testing data is generated by an unseen vocoder).

N\"uller et al. \cite{muller2022does} provide a comprehensive evaluation of audio deepfake detectors' generalization capabilities. They found that current detection methods, trained or designed based on the ASVspoof benchmark, perform poorly when detecting novel real-world audio deepfakes.
To date, limited research has addressed the issue of generalization in detecting audio deepfakes. Previous efforts have focused on 
domain-invariant representation learning \cite{xie2023domain} and self-supervised learning \cite{wang2023can}. However, these approaches rely on predefined vocoders and are often influenced by extraneous factors like background noise and speaker identity.

To tackle these challenges, we propose a novel method that improves the generalization of AI-synthesized voice detection by addressing both feature-level and optimization-level factors (see Figure \ref{fig:intro-i}). We start by experimentally analyzing the complex interactions of features and how the sharpness of loss landscapes impacts generalization in AI-synthesized voice detection.
We then introduce a novel disentanglement framework that combines multi-task and contrastive learning to extract domain-agnostic artifact features common across various vocoders. This process involves separating domain-agnostic from domain-specific artifacts (\ie, artifacts linked to specific vocoders) and applying reconstruction regularization to maintain consistency between the original and reconstructed voices.
To make the domain-agnostic features universally applicable and enhance generalization, we use content feature distribution as a benchmark and apply mutual information loss. This aligns the domain-agnostic features with the reference distribution.
Finally, we optimize the model by flattening the loss landscape to avoid suboptimal solutions and further improve generalization.
Our key contributions include:
\begin{enumerate}
    \item 
    We empirically analyze the factors from the feature and optimization levels affecting the generalization of AI-synthesized voice detection models.
    \item 
    We propose the first disentanglement framework to improve the generalization of AI-synthesized voice detection, focusing on both feature and optimization aspects. Specifically, we use disentanglement learning to extract domain-agnostic artifact features, which are utilized to enhance learning within a flattened loss landscape.
    \item Our extensive experiments conducted on various prominent audio deepfake datasets demonstrate the effectiveness of our framework, which surpasses the performance of state-of-the-art methods in improving the generalization for cross-domain detection. 
\end{enumerate}

\begin{figure*}
    \centering    \includegraphics[width=0.9\linewidth]{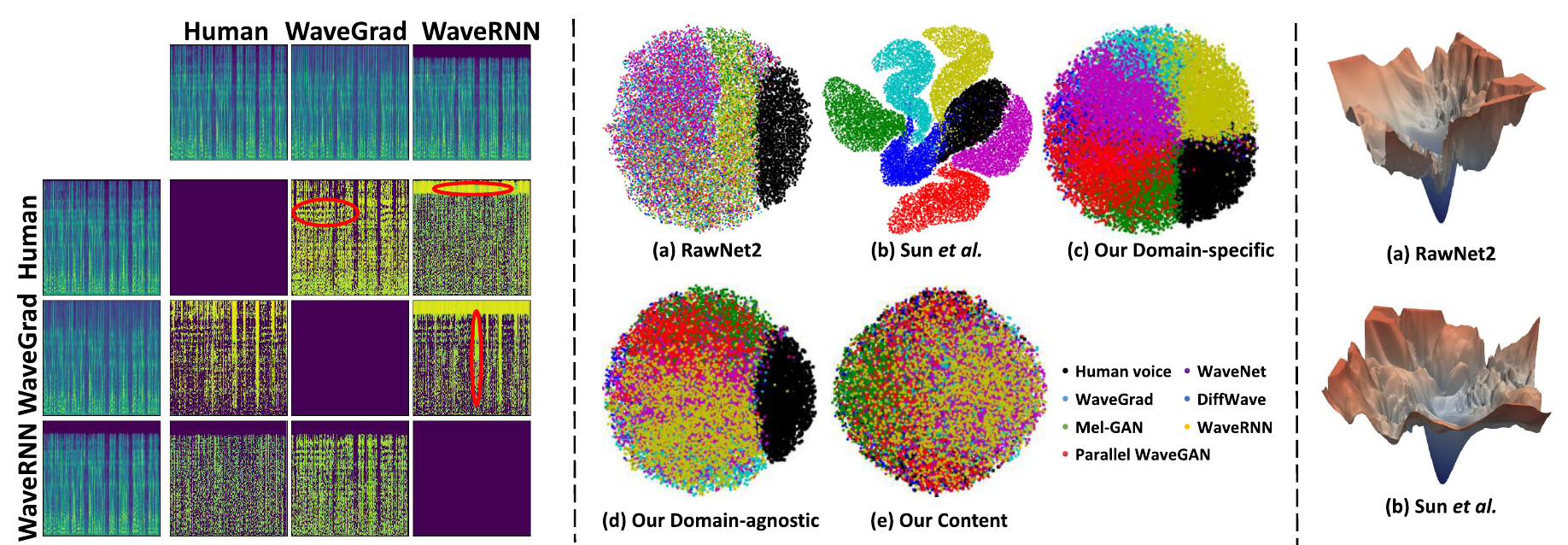}
    \caption{{Experimental results for Motivation. (\textbf{Left}) The differences of mel-spectrogram between human voices and AI-synthesized ones (\eg, based on WaveGrad~\cite{chen2020wavegrad} and WaveRNN~\cite{kalchbrenner2018efficient} vocoders). 
    The red circles highlight AI vocoder artifacts.
    More details about these differences can be found in Appendix. (\textbf{Middle}) The UMAP~\cite{2018arXivUMAP} visualization of features from related methods and our framework on LibriSeVoc~\cite{sun2023ai}. 
    The genuine and forged voices from six vocoders are separated in the latent space.
    The baselines (RawNet2 \cite{tak2021end}, Sun \emph{et al.}~\cite{sun2023ai}) and our domain-specific module can learn the domain-specific features, whereas the domain-agnostic module of our method captures the shared forgery features across different vocoders, and the content module exclusively captures forgery-irrelevant features. (\textbf{Right}) Visualization of loss landscape for RawNet2 and Sun \emph{et al}. The sharp local and global minima could lead to models with poor generalization.}
    }
    \label{fig:introduction-overall}
  \end{figure*}

\section{Related Work}

\textbf{Voice Synthesis}.
Recent years witnessed remarkable improvements in synthesized speech quality, driven by deep learning~\cite{tan2021survey}. Two main approaches, Text-to-Speech (TTS) and Voice Conversion (VC),
both utilize vocoders to synthesize voices.
Early neural vocoders, like WaveNet~\cite{oord2016wavenet} and WaveRNN~\cite{kalchbrenner2018efficient}, used features directly for waveform generation. Recent ones, such as SC-WaveRNN~\cite{paul2020scwavernn} and MB WaveRNN~\cite{chengzhu2020durian} adopted mel-spectrograms as input, but incurred slower inference times.
To address this, alternative generative methods have emerged in waveform generation. These include \textit{flow-based} methods like WaveGlow~\cite{prenger2019waveglow}, FloWaveNet~\cite{kim2019flowavenet}, and SqueezeWave~\cite{zhai2020squeezewave}. \textit{GAN-based} approaches namely MelGAN~\cite{kumar2019melgan}, Parallel WaveGAN~\cite{yamamoto2020parallel}, and Fre-GAN~\cite{lee2022fregan2}. \textit{VAE-based} techniques such as Wave-VAE~\cite{peng2020natts}. Additionally, \textit{diffusion-based} methods like WaveGrad~\cite{chen2020wavegrad}, DiffWave~\cite{kong2020diffwave}, and AudioLDM~\cite{liu2023audioldm}.

\textbf{AI-Synthesized Voice Detection}. 
AI-synthesized voice detection is a fundamental task in machine learning, distinguishing authentic human speech from synthesized audio. The end-to-end approaches like RawNet2~\cite{tak2021end} and RawFormer~\cite{liu2023rawformer} gained traction due to their competitive performance and efficiency. 
While these methods excel on intra-domain datasets, they often falter when dealing with cross-domain datasets. 
Some techniques \cite{salvi2023reliability, wang2023cross} 
Enhance generalization by identifying the most relevant signal segments or combining multi-view features to detect fake audio.
However, they rely on predefined vocoders and consider the entire feature space, making them vulnerable to background noise and speaker identity. These challenges are the focal points of our work.

\textbf{Disentangled Representation Learning}. 
Disentanglement representation learning \cite{wang2022disentangled} is used in voice conversion, and voice style transfer during speech synthesis~\cite{zhang2019learning, aloufi2020privacy,luong2021many, champion2022disentangled}. Only~\citet{yadav2023dsvae} applies disentanglement representation learning in AI-synthesized voice detection. However, their approach—which extracts features directly from speech spectrograms—does not account for the artifact features characteristic of different vocoders. Additionally, it fails to address the domain-specific features that are intrinsic to different synthesis techniques. This leads to a limited generalization capability since it only eliminates the influence of the content features but may still overfit domain-specific patterns. 
In contrast, our proposed method effectively disentangles both domain-specific and domain-agnostic artifact features associated with various vocoders.


\section{Motivating Experiments}\label{sec:motivation}

\textbf{Complex Entangled Information}.
The generalization issue in AI-synthesized voice detection arises from two main factors. 
Firstly, many detectors overly emphasize irrelevant content, like identity and background noise.
Secondly, different forgery techniques produce unique artifacts, as shown in Figure~\ref{fig:introduction-overall} (Left).
The red circle shows unique large-scale artifacts in various AI-synthesized voices, easily detected by a vocoder-specific detector.
However, detectors may become overly specialized in specific forgery, hindering generalization to unseen forgeries.
To support this hypothesis, we extract features from the LibriSeVoc~\cite{sun2023ai} dataset using RawNet2~\cite{tak2021end} and Sun \emph{et al.} \cite{sun2023ai}.
The UMAP visualizations \cite{2018arXivUMAP} show forgery vocoders' data clustering closely within baseline feature distribution (Figure \ref{fig:introduction-overall} Middle (a) and (b)), while different vocoders' data exhibit more distinct separations.
This phenomenon is also observed in the domain-specific feature distribution from our method (Figure \ref{fig:introduction-overall} (Middle (c)). 
Moreover, generalizable detectors should treat domain features from AI-synthesized voices equally but distinguish them from features of human voices (as illustrated in Figure \ref{fig:introduction-overall} (Middle (d))).
They should also treat content features equally, whether they are from human voices or AI-synthesized ones (as shown in Figure \ref{fig:introduction-overall} (Middle (e))).

\textbf{Sharpness of Loss Landscape}.
Existing DNN-based AI-synthesized voice detection models, such as RawNet2~\cite{tak2021end} and Sun \emph{et al} \cite{sun2023ai}, are highly overparameterized and tend to memorize data patterns during training. This results in sharp loss landscapes with multiple minima (Figure\ref{fig:introduction-overall} Right).
Such sharpness presents challenges for models to locate the correct global minima for better generalization. Flattening the loss landscape is crucial to smooth the optimization path and enable robust generalization.


\begin{figure*}[t]
  \centering
    \centerline{\includegraphics[width=1.0\linewidth]{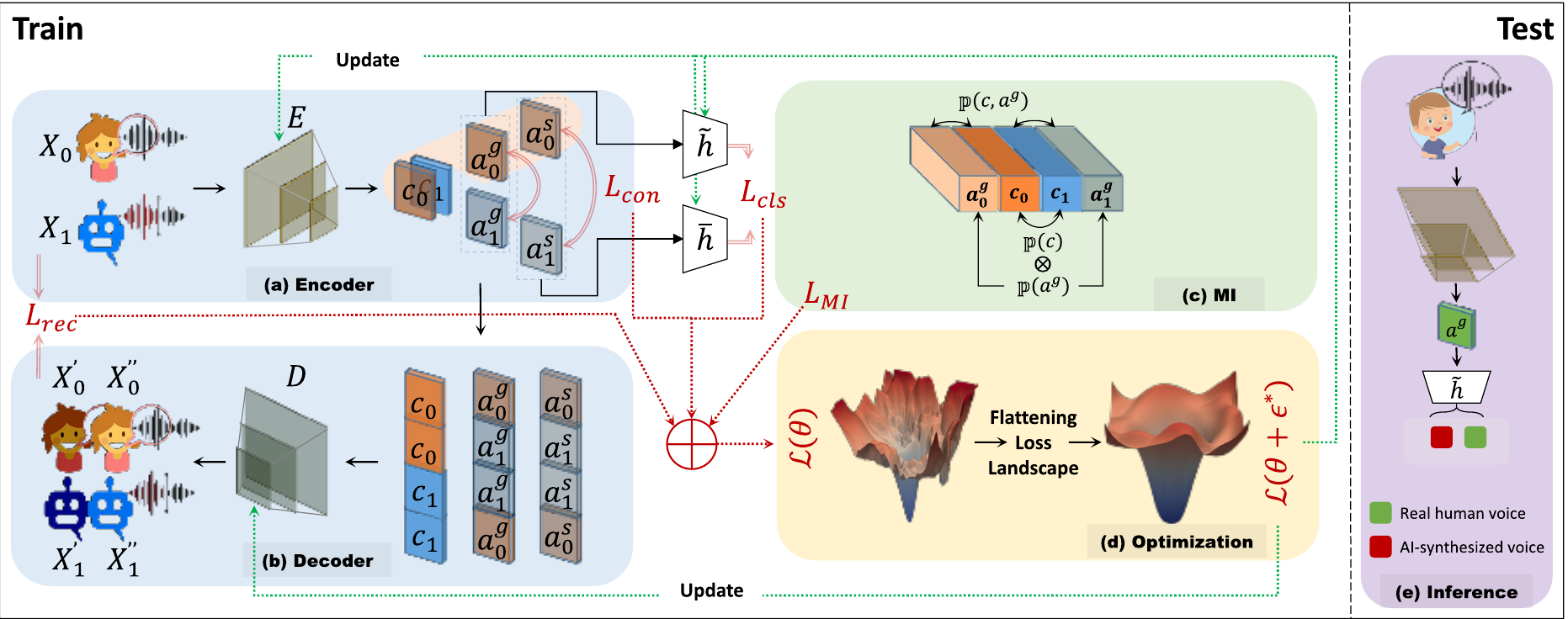}}
  \caption{{The overall architecture of our proposed approach. 
  (a) In the encoder module, two RawNet2~\cite{tak2021end}-structured backbones are added to the audio data signal to extract content and artifact features. Additionally, two headers further categorize artifact features into domain-agnostic and domain-specific categories.
  (b) In the decoder module, we use the content features as the base and fuse them with their own forgery features, as well as those from other samples for audio reconstruction.
  (c) Domain-agnostic features are made universally applicable by maximizing mutual information between domain-agnostic features and content features through joint and marginal distributions.
  (d) The sharpness-aware minimization (SAM) serves as an optimization technique to guide the model toward a flatter loss landscape to enhance its generalization.
  (e) For the inference, we take the predicted results of the domain-agnostic classification header. 
  }
  } 
  \label{fig:framework}
 \end{figure*}

\section{Method}

\textbf{Problem Setup}. Given a training dataset $\mathcal{S}=\{(X_i, D_i, Y_i)\}_{i=1}^n$ of size $n$, where $X_i \in \mathbb{R}^d$ represents the waveform of voice with feature dimension $d$ and $Y_i$ corresponds to the target label (\eg, synthetic voice or real voice). 
Moreover, $D_i$ serves as the domain label, indicating the voice generation source of $X_i$. For instance, in the LibriSeVoc dataset {\cite{sun2023ai}}, $D_i\in\{$\text{real}, \text{WaveNet} \cite{oord2016wavenet}, \text{WaveRNN} \cite{kalchbrenner2018efficient}, \text{MelGAN} \cite{kumar2019melgan}, \text{Parallel WaveGAN} \cite{yamamoto2020parallel}, \text{WaveGrad} \cite{chen2020wavegrad}, \text{DiffWave} \cite{kong2020diffwave}$\}$
Our goal is to train a synthesized voice detector on $\mathcal{S}$ for promising performance and high generalization to unseen synthesized voice data.

\textbf{Framework}. Our method, shown in Figure \ref{fig:framework}, consists of three main components: an encoder, a decoder, and two classification heads.
The encoder contains two parts: a content encoder and an artifact encoder, each responsible for extracting content and artifact features.
These two encoders share a common structure but operate with distinct sets of parameters.
The decoder facilitates audio reconstruction by utilizing artifact features as a conditioning factor alongside content features.
The classification module uses two heads: one for capturing domain-specific features and the other for extracting domain-agnostic features across different vocoders.
The training process is guided by an optimization module designed to flatten the loss landscape.
More details about the architecture can be found in Appendix.

\subsection{Disentanglement Process} 
We introduce a disentanglement learning module to extract vocoder-agnostic artifact features from the input voice for detection.
To elaborate, consider a pair of voices ($X_i$, $X_{j}$), where $i,j\in\{1,\cdots,n\}$ and $i\neq j$.
Here, $X_i$ represents the synthetic human voice (or real human voice), and $X_j$ corresponds to a real human voice (or synthetic human voice). Our encoder, denoted as $\mathbf{E}(\cdot)$, comprises both a content encoder and an artifact encoder for the extraction of content features $c$ and artifact features $a$.  
Notably, artifact features include domain-specific artifacts $a^s$ (vocoder-specific features) and domain-agnostic artifacts $a^g$ (common features across different vocoder models).
The operation of the encoder can be represented as follows: $c_i, a_i^s, a_i^g=\mathbf{E}(X_i)$.

\textbf{Classification Loss}.
To separate domain-specific artifacts from domain-agnostic ones, we use a simple approach with multi-task learning, applying cross-entropy loss to each. The loss function can be formulated as follows:
\( L_{cls} = C(\overline{h}(a_i^s), D_i) + \lambda_1 C(\widetilde{h}(a_i^g), Y_i) \), where $C(\cdot,\cdot)$ represents the CE loss. $\overline{h}$ and $\widetilde{h}$ denote the classification heads for $a_i^s$ and $a_i^g$, respectively. $\lambda_1$ represents a hyperparameter. 
Training with this classification loss allows the encoder to learn both specific and shared artifact information, improving the model's generalization capabilities.

\textbf{Contrastive Loss}.
The classification loss mentioned above only considers individual voice information, neglecting the important global correlations between voices that can improve the encoder's representation. Inspired by contrastive learning \cite{oord2018representation, yan2023ucf, lin2024preserving}, we address this by introducing a contrastive loss:
\( L_{con} = [b + \|a_{\text{anchor}} - a_+\|_2 - \|a_{\text{anchor}} - a_-\|_2]_+ \), where $a_{\text{anchor}}$ represents anchor artifact features of a voice, and $a_+$ and $a_-$ correspond to its positive counterpart from the same source and its negative counterpart from a different source, respectively. The hyperparameter $b$ is introduced, and $[\cdot]_+=\max\{0,\cdot\}$ denotes a hinge function.  
We apply $L_{con}$ to both domain-specific and domain-agnostic artifact features. For domain-specific features, the source is the neural vocoder type, and the contrastive loss drives the encoder to capture specific vocoder representations. For domain-agnostic features, the source is either synthetic or real human voice, encouraging the encoder to learn a generalizable representation independent of any particular vocoder.

\textbf{Reconstruction Loss}.
To ensure the integrity of the extracted features and maintain consistency between the original and reconstructed voices, we apply a reconstruction loss, defined as follows:
\( L_{rec} = \|X_i - \mathbf{D}(c_i, a^s_i, a^g_i)\|_1 + \|X_i - \mathbf{D}(c_i, a^s_j, a^g_j)\|_1 \), where $\mathbf{D}(\cdot,\cdot,\cdot)$ represents a decoder responsible for voice reconstruction based on the disentangled feature representations. 
In the $L_{rec}$ loss, the first term is the self-reconstruction loss, which uses the input voice's latent features to minimize reconstruction errors. The second term is the cross-reconstruction loss, which penalizes errors using the partner's forgery feature. Together, these terms promote feature disentanglement.

\begin{table*}[ht!]
    \centering
    \caption{{Comparisons of intra (seen)/cross (unseen)-domain generalization ability with comparable methods under EER (\%). `LSV', `ASP', `WF', and `FAVC' stand for LibriSeVoc, ASVspoof2019, WaveFake, and FakeAVCeleb, respectively. `Avg' represents the average EER (\%) across the seen or unseen vocoders within each dataset.}}
    \scalebox{0.7}{
        \begin{tabular}{c|cccccccc|cccccccc}
        \hline
        \multirow{3}{*}{Methods} & \multicolumn{8}{c|}{LibriSeVoc} & \multicolumn{8}{c}{ASVspoof2019} \\ \cline{2-17} 
         & \multicolumn{4}{c|}{Seen vocoder} & \multicolumn{4}{c|}{Unseen vocoder} & \multicolumn{3}{c|}{Seen vocoder} & \multicolumn{5}{c}{Unseen vocoder} \\ \cline{2-17} 
         & \multicolumn{1}{c|}{Avg} & LSV & ASP & \multicolumn{1}{c|}{WF} & \multicolumn{1}{c|}{Avg} & ASP & FAVC& WF & \multicolumn{1}{c|}{Avg} & ASP & \multicolumn{1}{c|}{LSV}& \multicolumn{1}{c|}{Avg} & ASP & FAVC & LSV & WF \\ \hline
        LCNN~\cite{lavrentyeva2019stc} & \multicolumn{1}{c|}{34.04} & 7.80 & 46.21 & \multicolumn{1}{c|}{48.10} & \multicolumn{1}{c|}{45.08} & 41.90 & 49.28 & 44.06 & \multicolumn{1}{c|}{33.02} & 16.15 & \multicolumn{1}{c|}{49.88} & \multicolumn{1}{c|}{41.21} & 9.81 & 50.98 & 51.93 & \textbf{52.13} \\
        RawNet2~\cite{tak2021end} & \multicolumn{1}{c|}{20.21} & 1.59 & 29.86 & \multicolumn{1}{c|}{29.18} & \multicolumn{1}{c|}{30.16} & 24.09 & 33.92 & 32.47 & \multicolumn{1}{c|}{20.79} & 1.89 & \multicolumn{1}{c|}{39.68} & \multicolumn{1}{c|}{39.55} & 6.46 & 51.37 & 47.71 & 52.65 \\
        WavLM~\cite{chen2022wavlm} & \multicolumn{1}{c|}{27.26} & 14.12 & 32.88 & \multicolumn{1}{c|}{34.79} & \multicolumn{1}{c|}{29.54} & 27.18 & 25.64 & 35.80 & \multicolumn{1}{c|}{50.75} & 14.22 & \multicolumn{1}{c|}{87.28} & \multicolumn{1}{c|}{59.27} & 7.91 & 83.84 & 87.28 & 58.07 \\
        XLS-R~\cite{babu2021xls} & \multicolumn{1}{c|}{33.47} & 11.21 & 45.37 & \multicolumn{1}{c|}{43.83} & \multicolumn{1}{c|}{45.11} & 51.23 & 42.91 & 41.18 & \multicolumn{1}{c|}{53.65} & 9.04 & \multicolumn{1}{c|}{98.26} & \multicolumn{1}{c|}{74.85} & 6.40 & 94.91 & 98.26 & 99.82 \\
        Sun \emph{et al.}~\cite{sun2023ai} & \multicolumn{1}{r|}{18.67} & 3.79 & 22.77 & \multicolumn{1}{c|}{29.45} & \multicolumn{1}{r|}{27.86} & 24.42 & 25.52 & 33.65 & \multicolumn{1}{c|}{22.18} & 3.92 & \multicolumn{1}{c|}{40.44} & \multicolumn{1}{c|}{41.59} & 8.38 & 54.78 & 50.59 & 52.62 \\ \hline
        \baseline{\textbf{Ours}} & \multicolumn{1}{c|}{\baseline{\textbf{13.55}}} &{\baseline{\textbf{0.30}}} & {\baseline{\textbf{15.66}}} & \multicolumn{1}{c|}{\baseline{\textbf{24.69}}} & \multicolumn{1}{c|}{\baseline{\textbf{20.27}}} & {\baseline{\textbf{16.29}}} & {\baseline{\textbf{18.02}}} & \baseline{\textbf{26.50}} & \multicolumn{1}{c|}{\baseline{\textbf{20.39}}} & {\baseline{\textbf{1.55}}} & \multicolumn{1}{c|}{\baseline{\textbf{39.23}}} & \multicolumn{1}{c|}{\baseline{\textbf{38.20}}} & {\baseline{\textbf{5.72}}} &{\baseline{\textbf{48.43}}} & {\baseline{\textbf{46.42}}} & \baseline{52.24} \\ \hline
        \end{tabular}
    }
    
    \label{tab:abl-on-various-data}
\end{table*}

\textbf{Mutual Information Loss}.
To enhance generalization, it's crucial to maintain consistent distributions of domain-agnostic features across different vocoders. This can be directly achieved by aligning their mutual relationships with the distributions of content features.
While linear dependence/independence techniques \cite{he2017learning, he2018wasserstein} could be considered for this purpose, they often fail to capture the mutual relationships between content and domain-agnostic features in high-dimensional, nonlinear spaces. In contrast, mutual information \cite{kinney2014equitability} is more effective for capturing arbitrary dependencies between variables.

Inspired by \cite{belghazi2018mutual}, we employ the Kullback-Leibler (KL) divergence \cite{joyce2011kullback}, which is an equivalent form of mutual information, to quantify the dependencies between $c$ and $a^g$. This is expressed as follows: $\text{MI}(c; a^g) = \mathbb{D}_{\text{KL}}(\mathbb{P}(c,a^g)|| \mathbb{P}(c) \otimes \mathbb{P}(a^g)).$
In this equation, $\mathbb{P}(\cdot,\cdot)$ represents the joint probability distribution, $\mathbb{P}(\cdot)$ denotes the marginal probability distribution, $\otimes$ signifies the product of the marginals, and $\mathbb{D_{KL}}$ is the KL divergence.

Given that the probability densities $\mathbb{P}(c,a^g)$ and $\mathbb{P}(c) \otimes \mathbb{P}(a^g)$ are not directly known, {maximizing} $\mathbb{D}_{\text{KL}}(\mathbb{P}(c,a^g)|| \mathbb{P}(c) \otimes \mathbb{P}(a^g))$ is a challenging task. However, we can maximize its lower bound $L_{MI}$ using the Donsker-Varadhan representation \cite{donsker1983asymptotic}, which can be expressed as:
\begin{equation*}
\begin{aligned}
L_{MI}:=\mathbb{E}_{x\sim\mathbb{P}(c,a^g)}[T(x)] - \log \mathbb{E}_{x\sim\mathbb{P}(c) \otimes \mathbb{P}(a^g)}[e^{T(x)}],
\end{aligned}
\label{eq:mi}
\end{equation*}
{where $T: \mathbb{R}^{d_c}\times \mathbb{R}^{d_{a^g}} \rightarrow \mathbb{R}$ represents a mutual information estimator. Inspired by \cite{hjelm2018learning}, $T$ can be absorbed into the encoder, combining the content feature $c_i$ and domain-agnostic feature $a_i^g$ in practice.}

\subsection{Optimization with Flattening Loss Landscape}
To sum up, the final loss function can be expressed as
\begin{equation}
    \begin{aligned}        \min_{\theta} \mathcal{L}(\theta):=\frac{1}{n}\sum_i[L_{cls}+\lambda_2 L_{con}+\lambda_3 L_{rec}]- \lambda_4 L_{MI}.
    \end{aligned}
\label{eq:overall_loss}
\end{equation}
Here, we assume that the model weights of the entire framework are denoted as $\theta$. The hyperparameters $\lambda_2$, $\lambda_3$, and $\lambda_4$ are introduced to strike a balance between each term in the loss function.
In practice, Eq. (\ref{eq:overall_loss}) can be solved by using a gradient descent approach to update $\theta$.
To help the model avoid suboptimal solutions common in overparameterized DNNs and further improve generalization, we apply the SAM technique \cite{foret2020sharpness} to flatten the loss landscape.  
Specifically, this involves finding an optimal $\epsilon^*$ to perturb $\theta$ in a way that maximizes the loss, expressed as:
\begin{equation}\footnotesize
    \begin{aligned}        
    \epsilon^*=\arg\max_{\|\epsilon\|_2\leq \gamma} \mathcal{L}(\theta+\epsilon)\approx\arg\max_{\|\epsilon\|_2\leq \gamma} \epsilon^\top\nabla_\theta \mathcal{L}=\gamma\sign(\nabla_\theta \mathcal{L}),
    \end{aligned}
\label{eq:eq_sam}
\end{equation}
Here, $\gamma$ is a hyperparameter that controls the perturbation magnitude, and $\nabla_\theta \mathcal{L}$ is the gradient of $\mathcal{L}$ with respect to $\theta$. The approximation term is derived using a first-order Taylor expansion, assuming $\epsilon$ is small. The final equation results from solving a dual norm problem, using the $\text{sign}$ function. Thus, the model weights are updated by solving:
\begin{equation}
    \begin{aligned}        
    \min_\theta \mathcal{L}\textbf{(}\theta+\epsilon^*\textbf{)}.
    \end{aligned}
\label{eq:sharpness}
\end{equation}
The underlying idea is that perturbing the model in the direction of the gradient norm increases the loss value, thereby improving generalization.  We optimize Eq. (\ref{eq:sharpness}) using stochastic gradient descent, and the related algorithm is provided in the Appendix. Note that this is the first time to adapt SAM for AI-synthesized voice detection. Our ablation study demonstrates the effectiveness of this novel application of it in enhancing generalization. 

\section{Experiment}
\subsection{Experimental Settings}
\textbf{Datasets}.
To assess the generalization of our method, we tested it on various mainstream audio benchmarks, including LibriSeVoc~\cite{sun2023ai}, WaveFake~\cite{frank2021wavefake}, ASVspoof 2019~\cite{lavrentyeva2019stc}, and the audio segment of FakeAVCeleb~\cite{khalid2021fakeavceleb}.
More details of them and evaluations on other datasets (\eg, ASVspoof2021 {\cite{yamagishi2021asvspoof}}) are in Appendix.

\textbf{Baseline Methods}.
To assess our method's generalization capacity, we compared it with the following baselines:
\textbf{1)} \textbf{\textit{LCNN}}~\cite{lavrentyeva2019stc} achieved the second-best performance in ASVspoof 2021 Speech Deepfake track. \textbf{2)} \textbf{\textit{RawNet2}}~\cite{tak2021end} achieved the top performance in ASVspoof 2021 Speech Deepfake track. \textbf{3)} \textbf{\textit{WavLM}}~\cite{chen2022wavlm}, developed by Microsoft, is a multilingual pre-trained model for general audio tasks. 
\textbf{4)} \textbf{\textit{XLS-R}}~\cite{babu2021xls}, a robust cross-lingual speech model, excels in various domains, such as speech translation, speech recognition, and language identification.
\textbf{5)} \textbf{\textit{Sun {et al}.}}~\cite{sun2023ai} is the latest method in detecting AI-synthesized voices by identifying the artifacts of vocoders in audio signals. 

\textbf{Evaluation Metrics}.
We evaluate our method using the Equal Error Rate (EER) metric, as commonly employed in previous studies~\cite{frank2021wavefake,yamagishi2021asvspoof} and baselines. A lower value means better performance. {Furthermore, we add additional evaluations \cite{yamagishi2021asvspoof} that include metrics AUC, False Acceptance Rate of Synthesized Voices, Rejection Rate of Real Voices, and Detection Cost Function in the Appendix, which also show the strong generalization of our method.}

\textbf{Implementation Details.}
Our encoders are based on RawNet2~\cite{tak2021end} but exclude the last fully connected layer and utilize only features before it as encoder outputs. 
We employ the Adam~\cite{kingma2014adam} optimizer with a learning rate set to 0.0002 and a batch size of 16.
Hyperparameters $\lambda_1$, $\lambda_2$, $\lambda_3$, and $\lambda_4$ are set to 0.1, 0.3, 0.05, and 0.03, respectively. 
The margin $b$ in $L_{con}$ is set to $3$. The $\gamma$ in Eq. (\ref{eq:eq_sam}) is set to 0.07. 
We also use the original voice signal as input and apply the same data preprocessing as RawNet2~\cite{tak2021end}, padding all signals to the same size.
{More details can be found in the Appendix}.

\begin{table}[t]
\centering
\caption{{Detection EER (\%) of cross-dataset evaluated on WaveFake Benchmark and trained on LibriSeVoc.} 
}
\scalebox{0.88}{
    \tablestyle{1.5pt}{1.3}
    \begin{tabular}{cc|ccccc|c}
    \hline
    \multicolumn{2}{c|}{Vocoders} & \begin{tabular}[c]{@{}c@{}}LCNN\\ \end{tabular} & \begin{tabular}[c]{@{}c@{}}RawNet2\\ \end{tabular} & \begin{tabular}[c]{@{}c@{}}WavLM\\ \end{tabular} & \begin{tabular}[c]{@{}c@{}}XLS-R\\ \end{tabular} & \begin{tabular}[c]{@{}c@{}}Sun \emph{et al.}\\ \end{tabular} & \baseline{\textbf{Ours}}  \\ \hline
    \multicolumn{1}{c|}{\multirow{3}{*}{Seen}} & MelGAN & 26.77 & 9.60 & 37.10 & 47.08 & 17.16 & \baseline{\textbf{2.93}} \\
    \multicolumn{1}{c|}{} & PWGAN & 59.21 & 43.81 & 31.05 & 40.08 & 36.71 & \baseline{\textbf{31.12}}  \\ \cline{2-8} 
    \multicolumn{1}{c|}{} & Avg & 42.99 & 26.70 & 34.08 & 43.58 & 26.94 & \baseline{\textbf{17.03}}  \\ \hline
    \multicolumn{1}{c|}{\multirow{5}{*}{Unseen}} & WvGlow & 28.21 & 4.14 & 36.47 & 31.34 & 12.87 & \baseline{\textbf{0.50}} \\
    \multicolumn{1}{c|}{} & MBMGAN & 46.06 & 37.21 & 31.77 & 44.13 & 37.91 & \baseline{\textbf{22.92}} \\
    \multicolumn{1}{c|}{} & FBMGAN & 48.69 & 44.65 & 37.08 & 45.42 & 46.85 & \baseline{\textbf{39.79}} \\
    \multicolumn{1}{c|}{} & HiFiGAN & 38.98 & 38.58 & 37.69 & 40.69 & 38.60 & \baseline{\textbf{29.76}}\\ \cline{2-8} 
    \multicolumn{1}{c|}{} & Avg & 40.49 & 31.14 & 35.75 & 40.40 & 34.06 & \baseline{\textbf{23.38}} \\ \hline
    \end{tabular}
}
\label{tab:cross-vocoders}
\end{table}

\subsection{Results}
\textbf{Performance on Various Datasets}.
To demonstrate the universality of our method, we expand our training set to include data from LibriSeVoc~\cite{sun2023ai} and ASVspoof2019~\cite{lavrentyeva2019stc}, respectively. 
Our model is subsequently evaluated on four distinct datasets: LibriSeVoc~\cite{sun2023ai}, ASVspoof2019~\cite{lavrentyeva2019stc}, WaveFake~\cite{frank2021wavefake}, and FakeAVCeleb~\cite{khalid2021fakeavceleb}. 
We divide the test sets into two categories: seen vocoders from the same domain and unseen vocoders for cross-domain evaluation, based on the vocoder categories present in the training set. 
{More details of dataset-vocoder partitions can be found in Appendix}.

As shown in Table \ref{tab:abl-on-various-data}, our method excels baselines on the same domain and cross-domain scenarios. 
Specifically, our method outperforms Sun \emph{et al.}~\cite{sun2023ai} by nearly 5.12\% (13.55\% vs. 18.67\% trained on LibriSeVoc) and 1.79\% (20.39\% vs. 22.18\% trained on ASVspoof2019) in the intra-domain setting, and by approximately 7.59\% (20.27\% vs. 27.86\% trained on LibriSevoc) and 1.34\% (38.20\% vs. 39.55\% trained on ASVspoof2019) in the cross-domain scene.
Furthermore, despite the exceptional performance of the baselines in various audio downstream tasks, WavLM~\cite{chen2022wavlm} and XLS-R~\cite{babu2021xls} exhibit inferior performance in detecting AI-synthesized voices.
While the LCNN method excels in performance for vocoder `WF' on ASVspoof2019, our approach surpasses others in overall performance.

\textbf{Performance on Vocoders}. 
To assess our method on intra-domain and cross-domain vocoders, we train the model on LibriSeVoc~\cite{sun2023ai} and evaluate it on WaveFake~\cite{frank2021wavefake}. We categorize WaveFake into seen/unseen subsets and conduct testing on all vocoders within each subset.
The results are reported in Table \ref{tab:cross-vocoders}. We observe that our method not only achieves the best performance on seen vocoders but also exhibits a significant performance boost on unseen vocoders (23.38\% vs. 34.06\%).

\begin{table}[t]
\centering
\caption{{Comparisons of generalization ability with compared methods in EER (\%). We use LSV (LibriSeVoc) as a train set, and test on four datasets: LSV, WF, ASP, and FAVC. The abbreviations `w/o WaveNet' represent the voices generated by the WaveNet vocoder and are removed from LSV.}}
\scalebox{0.88}{
    \tablestyle{1.5pt}{1.3}
    \begin{tabular}{c|c|cccc|cccc}
    \hline
    \multirow{2}{*}{Train Set} & \multirow{2}{*}{Methods} & \multicolumn{4}{c|}{Seen vocoder} & \multicolumn{4}{c}{Unseen vocoder} \\ \cline{3-10} 
     &  & \multicolumn{1}{c|}{Avg} & LSV &ASP& \multicolumn{1}{c|}{WF} & \multicolumn{1}{c|}{Avg} & ASP &FAVC & \multicolumn{1}{c}{WF} \\ \hline
    \multirow{4}{*}{\parbox{2cm}{
        \centering
        \begin{tabular}{c}
        LSV \\
        w/o \\
        WaveNet
        \end{tabular}
        }
        }& LCNN & \multicolumn{1}{c|}{33.9} & 8.3 & 45.1 & 48.4 & \multicolumn{1}{c|}{46.2} & 40.6 & 54.1 & 44.0 \\
     & RawNet2 & \multicolumn{1}{c|}{21.2} & 7.5 & 36.5 & 19.6 & \multicolumn{1}{c|}{27.2} & 28.2 & \textbf{23.0} & 30.4 \\
     & Sun \emph{et al.} & \multicolumn{1}{c|}{23.9} & 8.6 & 36.9 & 26.3 & \multicolumn{1}{c|}{32.1} & 30.8 & 32.0 & 33.5 \\  \cline{2-10}
     & \baseline{\textbf{Ours}} & \multicolumn{1}{c|}{\baseline{\textbf{13.9}}} & \baseline{\textbf{1.9}} & \baseline{\textbf{20.5}} & \baseline{\textbf{19.2}} & \multicolumn{1}{c|}{\baseline{\textbf{22.7}}} & \baseline{\textbf{19.7}} & \baseline{23.3} & \baseline{\textbf{24.9}} \\ \hline
    \multirow{4}{*}{\parbox{2cm}{
        \centering
        \begin{tabular}{c}
        LSV \\
        w/o \\
        WaveRNN
        \end{tabular}}} & LCNN & \multicolumn{1}{c|}{36.5} & 9.8 & 49.9 & 50.0 & \multicolumn{1}{c|}{47.7} & 50.0 & 43.5 & 49.7 \\
     & RawNet2 & \multicolumn{1}{c|}{23.8} & 11.2 & 34.7 & 25.3 & \multicolumn{1}{c|}{29.5} & 30.2 & 30.7 & 27.5 \\
     & Sun \emph{et al.} & \multicolumn{1}{c|}{25.3} & 13.1 & 33.3 & 29.4 & \multicolumn{1}{c|}{28.7} & 23.1 & 30.8 & 32.1 \\ \cline{2-10}
     & \baseline{\textbf{Ours}} & \multicolumn{1}{c|}{\baseline{\textbf{18.6}}} & \baseline{\textbf{8.7}} & \baseline{\textbf{24.8}} & \baseline{\textbf{22.3}} & \multicolumn{1}{c|}{\baseline{\textbf{23.5}}} & \baseline{\textbf{22.1}} & \baseline{\textbf{22.7}} & \baseline{\textbf{25.6}} \\ \hline
    \multirow{4}{*}{\parbox{2cm}{
        \centering
        \begin{tabular}{c}
        LSV \\
        w/o \\
        WaveGrad
        \end{tabular}
        }} & LCNN & \multicolumn{1}{c|}{33.1} & 7.8 & 43.9 & 47.5 & \multicolumn{1}{c|}{44.6} & 42.1 & 45.8 & 45.9 \\
     & RawNet2 & \multicolumn{1}{c|}{18.4} & 2.8 & 24.9 & 27.4 & \multicolumn{1}{c|}{29.0} & 20.7 & 35.6 & 30.6 \\
     & Sun \emph{et al.} & \multicolumn{1}{c|}{24.8} & 3.0 & 34.7 & 36.6 & \multicolumn{1}{c|}{38.1} & 27.8 & 45.0 & 41.5 \\ \cline{2-10}
     & \baseline{\textbf{Ours}} & \multicolumn{1}{c|}{\baseline{\textbf{15.0}}} & \baseline{\textbf{1.1}} & \baseline{\textbf{20.1}} & \baseline{\textbf{23.9}} & \multicolumn{1}{c|}{\baseline{\textbf{23.3}}} & \baseline{\textbf{18.8}} & \baseline{\textbf{22.3}} & \baseline{\textbf{28.6}} \\ \hline
    \multirow{4}{*}{\parbox{2cm}{
        \centering
        \begin{tabular}{c}
        LSV \\
        w/o \\
        DiffWave
        \end{tabular}
        }} & LCNN & \multicolumn{1}{c|}{37.3} & 10.0 & 52.3 & 49.7 & \multicolumn{1}{c|}{50.0} & 49.0 & 56.2 & 45.0 \\
     & RawNet2 & \multicolumn{1}{c|}{24.5} & 2.6 & 39.5 & 31.4 & \multicolumn{1}{c|}{34.6} & 29.7 & 40.3 & 33.8 \\
     & Sun \emph{et al.} & \multicolumn{1}{c|}{29.3} & 6.7 & 38.7 & 42.4 & \multicolumn{1}{c|}{33.4} & 29.4 & 33.4 & 37.4 \\ \cline{2-10}
     & \baseline{\textbf{Ours}} & \multicolumn{1}{c|}{\baseline{\textbf{16.4}}} & \baseline{\textbf{1.2}} & \baseline{\textbf{17.3}} & \baseline{\textbf{30.6}} & \multicolumn{1}{c|}{\baseline{\textbf{27.5}}} & \baseline{\textbf{19.8}} & \baseline{\textbf{31.5}} & \baseline{\textbf{31.3}} \\ \hline
    \multirow{4}{*}{\parbox{2cm}{
        \centering
        \begin{tabular}{c}
        LSV \\
        w/o \\
        MelGAN
        \end{tabular}
        }
    } & LCNN & \multicolumn{1}{c|}{36.4} & 12.5 & 49.8 & 46.8 & \multicolumn{1}{c|}{45.6} & 48.2 & 47.6 & 41.1 \\
     & RawNet2 & \multicolumn{1}{c|}{21.2} & 1.0 & 36.1 & 26.4 & \multicolumn{1}{c|}{27.0} & 26.7 & 25.7 & 28.5 \\
     & Sun \emph{et al.} & \multicolumn{1}{c|}{21.4} & 0.8 & 35.6 & 27.8 & \multicolumn{1}{c|}{29.4} & 25.4 & 29.6 & 33.2 \\ \cline{2-10}
     & \baseline{\textbf{Ours}} & \multicolumn{1}{c|}{\baseline{\textbf{13.6}}} & \baseline{\textbf{0.8}} & \baseline{\textbf{17.7}} & \baseline{\textbf{22.3}} & \multicolumn{1}{c|}{\baseline{\textbf{24.1}}} & \baseline{\textbf{20.3}} & \baseline{\textbf{25.4}} & \baseline{\textbf{26.6}} \\ \hline
    \multirow{4}{*}{\parbox{2cm}{\centering
        \begin{tabular}{c}
        LSV \\
        w/o \\
        PWaveGan
        \end{tabular}}} & LCNN & \multicolumn{1}{c|}{36.0} & 7.9 & 50.0 & 50.3 & \multicolumn{1}{c|}{47.0} & 46.5 & 50.0 & 44.6 \\
     & RawNet2 & \multicolumn{1}{c|}{21.1} & 2.1 & 32.7 & 28.3 & \multicolumn{1}{c|}{28.8} & 26.4 & 27.3 & 32.8 \\
     & Sun \emph{et al.} & \multicolumn{1}{c|}{27.9} & 3.9 & 43.8 & 36.0 & \multicolumn{1}{c|}{31.2} & 32.3 & 27.7 & 33.6 \\ \cline{2-10}
     & \baseline{\textbf{Ours}} & \multicolumn{1}{c|}{\baseline{\textbf{16.8}}} & \baseline{\textbf{0.8}} & \baseline{\textbf{26.0}} & \baseline{\textbf{23.5}} & \multicolumn{1}{c|}{\baseline{\textbf{26.2}}} & \baseline{\textbf{21.3}} & \baseline{\textbf{25.9}} & \baseline{\textbf{31.3}} \\ \hline
    \end{tabular}
}
\label{tab:generalization-on-data}
\end{table}


\textbf{Performance Affected by Specific Vocoder}. 
To further investigate the model's generalization capabilities and assess whether the models heavily rely on a specific vocoder, we remove one vocoder from LibriSeVoc~\cite{sun2023ai} one by one, resulting in six sub-training sets. 
Then, we conduct tests on LibriSeVoc~\cite{sun2023ai}, ASVspoof2019~\cite{lavrentyeva2019stc}, WaveFake~\cite{frank2021wavefake}, and FakeAVCeleb~\cite{khalid2021fakeavceleb}. 
Similarly, we categorize these benchmarks into seen and unseen domains and report the Equal Error Rate (EER) metric within these domains.
As shown in Table \ref{tab:generalization-on-data}, we observe that our method consistently achieves top performance on all six sub-training sets. 
Notably, Our feature decoupling approach generally performs better in unseen domains, demonstrating stable generalization in cross-domain scenarios.

\begin{table*}[]
    \centering
    \caption{{An analysis study of our key modules: RawNet2~\cite{tak2021end}, `Rec'(Reconstruction module), `Cls'(multi-task header), `Con'(contrastive learning module), `MI'({mutual information module}), and `SAM'. These modules are constructed and trained under LSV and then tested on seen/unseen parts of LSV, ASP, WF, and FAVC.} }
    \scalebox{1}{
    \tablestyle{6pt}{1.1}
        \begin{tabular}{c|cccccc|cccc|cccc}
        \hline
        \multirow{2}{*}{Method} & \multicolumn{6}{c|}{Module} & \multicolumn{4}{c|}{Seen vocoder} & \multicolumn{4}{c}{Unseen vocoder} \\ \cline{2-15} 
         & RawNet2 & Rec & Cls & Con & MI & SAM & \multicolumn{1}{c|}{Avg} & \multicolumn{1}{c}{LSV} & \multicolumn{1}{c}{ASP} & WF & \multicolumn{1}{c|}{Avg} & \multicolumn{1}{c}{ASP} & \multicolumn{1}{c}{FAVC} & WF \\ \hline
        RawNet2 & \checkmark &  &  &  &  &  & \multicolumn{1}{c|}{20.21} & \multicolumn{1}{c}{1.59} & \multicolumn{1}{c}{29.86} & 29.18 & \multicolumn{1}{c|}{30.16} & \multicolumn{1}{c}{24.09} & \multicolumn{1}{c}{33.92} & 32.47 \\ \hline
        $V_A$ & \checkmark & \checkmark &  &  &  &  & \multicolumn{1}{c|}{19.16} & \multicolumn{1}{c}{0.53} & \multicolumn{1}{c}{31.84} & 25.11 & \multicolumn{1}{c|}{28.57} & \multicolumn{1}{c}{27.49} & \multicolumn{1}{c}{26.35} & 31.87 \\
        $V_B$ & \checkmark & \checkmark & \checkmark &  &  &  & \multicolumn{1}{c|}{16.58} & \multicolumn{1}{c}{0.30} & \multicolumn{1}{c}{25.15} & 24.30 & \multicolumn{1}{c|}{26.07} & \multicolumn{1}{c}{22.38} & \multicolumn{1}{c}{27.13} & 28.71 \\
        $V_C$ & \checkmark & \checkmark & \checkmark & \checkmark &  &  & \multicolumn{1}{c|}{15.90} & \multicolumn{1}{c}{0.42} & \multicolumn{1}{c}{22.57} & 24.72 & \multicolumn{1}{c|}{26.83} & \multicolumn{1}{c}{24.15} & \multicolumn{1}{c}{27.64} & 28.71 \\
        $V_D$ & \checkmark & \checkmark & \checkmark & \checkmark & \checkmark &  & \multicolumn{1}{c|}{15.71} & \multicolumn{1}{c}{0.38} & \multicolumn{1}{c}{25.62} & \textbf{21.12} & \multicolumn{1}{c|}{23.23} & \multicolumn{1}{c}{23.52} & \multicolumn{1}{c}{20.79} & \textbf{25.37} \\ \hline
        \baseline{ \textbf{Ours}} & \baseline{\checkmark} & \baseline{\checkmark} & \baseline{\checkmark} & \baseline{\checkmark} & \baseline{\checkmark} & \baseline{\checkmark} & \multicolumn{1}{c|}{\baseline{\textbf{13.55}}} & \multicolumn{1}{c}{\baseline{\textbf{0.30}}} & \multicolumn{1}{c}{\baseline{\textbf{15.66}}} & \baseline{24.69} & \multicolumn{1}{c|}{\baseline{\textbf{20.27}}} & \multicolumn{1}{c}{\baseline{\textbf{16.29}}} & \multicolumn{1}{c}{\baseline{\textbf{18.02}}} & \baseline{26.50} \\ \hline
        \end{tabular}
      }
    \label{tab:abl-modules}
\end{table*}

\begin{figure*}[h]
    \centering\includegraphics[width=1.0\linewidth]{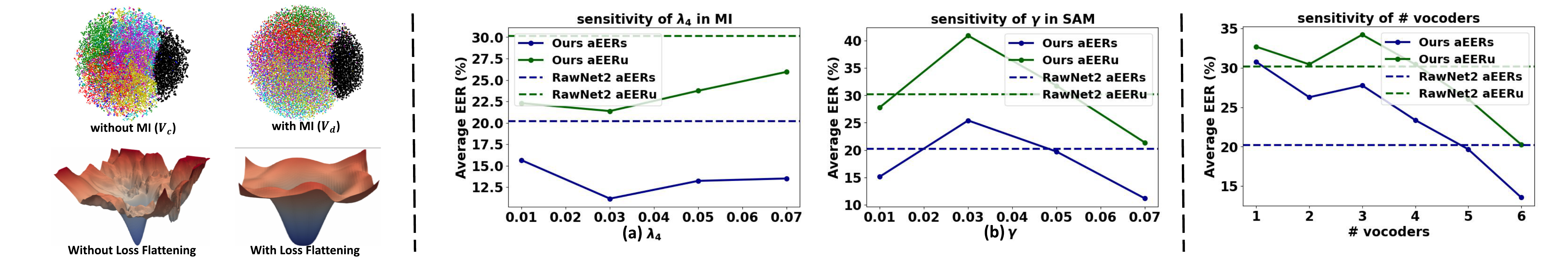}
    \caption{{(\textbf{Left-Top}) The virtualizations of model with MI ($V_c$) and model without MI ($V_d$). 
    (\textbf{Left-Bottom}) The loss landscape visualization of our method with and without flattening the loss landscape.
    (\textbf{Middle}) (a) The effect of $\lambda_4$ for balancing mutual information term. (b) The effect of $ \gamma $ in sharpness-aware minimization. (\textbf{Right}) The effect of the number of vocoders used in the train set. The reported average Equal Error Rate for seen (aEERs) and average Equal Error Rate for unseen (aEERu) domains.}}
    \label{fig:abl_sensitivity_all}
\end{figure*}

\subsection{Ablation Study}
\textbf{Analyzing the Framework Components in a Deconstructive Trajectory}. To evaluate the impact of each component in our proposed method, we conduct an analysis on several datasets. The results are reported in Table~\ref{tab:abl-modules}. 
Our modules consistently improve performance on seen or unseen vocoders.
More detailed observations are as follows: \textbf{1)} $V_A$ outperforms the baseline RawNet2~\cite{tak2021end} on most datasets, especially on seen WF and unseen FAVC, revealing the efficacy of our reconstruction module. 
\textbf{2)} $V_B$ and $V_C$ achieve relatively similar results and further enhance the performance compared with $V_A$ (\eg, 6.69\% and 9.27\% improvement of EER on seen ASP), indicating the necessity of our multi-task header and contrastive learning module. 
{\textbf{3)} Implementing the mutual information module ($V_D$) greatly improves the performance in cross-domain evaluation (\eg, 6.85\% enhancement of EER on unseen FAVC compared with $V_C$), 
To further illustrate the impact of mutual information, we present demographic-agnostic features in Figure \ref{fig:abl_sensitivity_all} (Left-Top). It clearly shows that incorporating the mutual information module in the model leads to a more irregular distribution of fake features in the feature space, independent of the forgery techniques employed.
}
\textbf{4)} Combining all components (Ours) achieves optimal results in both intra-domain and cross-domain evaluation, indicating the efficacy of the SAM approach. 
Figure \ref{fig:abl_sensitivity_all} (Left-Bottom) illustrates SAM's impact on the loss landscape, showing its ability to transform the rugged surface into a smoother one and enhance the model's generalization capability.

\textbf{Effect of Hyperparameters}. With an increasing impact of mutual information ($\lambda_4$), 
the model's performance improves on both seen and unseen data. 
However, as $\lambda_4$ continues to increase, the voice detection information within domain-agnostic features gradually diminishes, leading to a decline in the model's ability to distinguish between Human and AI-synthesized voices. 
With increasing $\lambda_4$, voice detection information in domain-agnostic features diminishes, reducing the model's ability to distinguish between human and AI-synthesized voices.
Notably, we observe optimal performance when $\lambda_4 = 0.03$ (Figure \ref{fig:abl_sensitivity_all} (Middle (a))).
As SAM intensity increases, model performance initially slightly decreases, followed by continuous improvement, and reaching optimal performance at $\gamma = 0.07$ (Figure \ref{fig:abl_sensitivity_all} (Middle (b))). This suggests that the model generalization benefits from a smoother loss landscape.

\textbf{Ablation on the number of vocoders in the train set}. 
To illustrate the impact of vocoder diversity in the training dataset on the model generalization. We create subsets of the training data with different combinations of vocoder types, ranging from 1 to 6, sourced from LibriSeVoc. 
Trained models are evaluated on the similar seen/unseen manner, and the aEERs and aEERu are reported. 
Figure \ref{fig:abl_sensitivity_all} (Right) indicates a clear trend: increasing vocoder diversity in training data enhances model generalizability. 

\section{Conclusion}
Existing AI-synthesized voice detection models are limited by predefined vocoders and sensitivity to factors like background noise and speaker identity.
While excelling in intra-domain evaluation, they struggle to generalize across different domains with emerging voice generation advancements.
To address this challenge, we propose a new disentanglement framework with a module extracting domain-agnostic features related to vocoders.
Furthermore, we use these features to aid model learning in a flattened loss landscape, helping the model escape suboptimal solutions and enhance generalization.
Extensive experiments on various datasets, compared to state-of-the-art methods, demonstrate the superior detection generalization of our framework.
\textbf{Limitation}: One limitation of our method is its dependency on datasets including AI-synthesized voices generated by different types of vocoders. A few datasets contain synthesized voice from a single vocoder, limiting the extraction of domain-agnostic artifact features.
\textbf{Future Work}: In future work, we aim to explore methods that can directly detect synthetic audio and improve detection generalization not limited to datasets containing multiple vocoders. We also plan to extend our approach to multi-modal tasks for audio and video detection.

\section{Acknowledgments}
This work is supported by the U.S. National Science Foundation (NSF) under grant IIS-2434967 and the National Artificial Intelligence Research Resource (NAIRR) Pilot and TACC Lonestar6.
The views, opinions and/or findings expressed are those of the author and should not be interpreted as representing the official views or policies of NSF and NAIRR Pilot.

\bibliography{aaai25}

\appendix
\newpage

\twocolumn[
\begin{@twocolumnfalse}
\begin{center}
\textbf{\LARGE Appendix}
\end{center}
\end{@twocolumnfalse}
]

\section{Related Work}
\label{apx:related_work}

\subsection{Voice Synthesis}
Over recent years, the advancement of deep learning and artificial intelligence has yielded remarkable enhancements in the quality of synthesized speech~\cite{tan2021survey}. 
Neural network-based text-to-speech (TTS) architectures comprise several essential components, namely, text analysis, acoustic models, and vocoders. The text analysis module serves as the initial stage, responsible for text preprocessing and the extraction of linguistic features. Subsequently, these linguistic features are employed to train acoustic models, which predict acoustic attributes such as spectrum and cepstrum. Finally, vocoders are employed to transform the predicted acoustic features into synthesized speech. 
In addition to these components, fully end-to-end TTS models have been developed, including notable works such as Char2Wav~\cite{sotelo2017char2wav}, ClariNet~\cite{ping2019clarinet}, FastSpeech 2s~\cite{ren2021fastspeech2}, EATS~\cite{donahue2021eats}, VITS~\cite{kim2021vits}, Wave-Tacotron~\cite{weiss2021wavetacotron}, and EfficientTTS~\cite{miao2021efficienttts}.

Voice conversion (VC) constitutes another pivotal facet of voice synthesis. VC is a process that seamlessly transposes the distinctive identity of one speaker onto another while preserving the speech's content. A typical VC procedure involves two key phases: the first phase entails voice analysis and decomposition, extracting individual components and characteristics. The second phase involves mapping and merging these extracted elements through reconstructions facilitated by a vocoder. 
Recent VC models predominantly operate within the mel-spectrum domain, employing deep neural network architectures to map mel-spectrograms to audio signals. These models adopt generative approaches such as using VAE~\cite{long2022vae,lian2022vae} and GAN~\cite{li2021starganv2,chen2022efficient}, to retrieve the speech elements within the input voice and harmoniously blend them with the stylistic attributes of the voice. Subsequently, the resulting mel-spectrogram is reconstructed into an audio waveform by leveraging a neural vocoder.
Vocoders serve as a crucial role in the field of both TTS and VC by acting as the final stage for audio waveform synthesis. In the early days, neural vocoders, such as WaveNet~\cite{oord2016wavenet}, Char2Wav~\cite{sotelo2017char2wav}, and WaveRNN~\cite{kalchbrenner2018efficient}, directly employed linguistic features as their model input to generate the audio waveform. Subsequent advancements introduced newer neural vocoders, such as Univ. WaveRNN~\cite{lorenzo2019uni}, SC-WaveRNN~\cite{paul2020scwavernn}, and MB WaveRNN~\cite{chengzhu2020durian}, which adopted mel-spectrograms as input to generate the audio waveform. These methods implemented autoregressive modeling to predict the distribution of each audio waveform sample based on all preceding samples, which often incurred prolonged inference times. To mitigate this challenge, alternative generative approaches have been explored in waveform generation. These include flow-based methods, exemplified by WaveGlow~\cite{prenger2019waveglow}, FloWaveNet~\cite{kim2019flowavenet}, and SqueezeWave~\cite{zhai2020squeezewave}; GAN based methods, such as MelGAN~\cite{kumar2019melgan}, Parallel WaveGAN~\cite{yamamoto2020parallel}, and Fre-GAN~\cite{lee2022fregan2}; VAE based techniques, as seen in Wave-VAE~\cite{peng2020natts}; and diffusion-based methods, illustrated by WaveGrad~\cite{chen2020wavegrad}, DiffWave~\cite{kong2020diffwave}, and AudioLDM~\cite{liu2023audioldm}.

\subsection{AI-Synthesized Voice Detection}
AI-synthesized voice detection is a fundamental task in machine learning, dedicated to distinguishing authentic human speech from artificially synthesized audio. Current research in voice detection can be broadly classified into two predominant approaches: the two-stage methodology and the end-to-end framework. Within the two-stage approach, the initial step involves the extraction of discriminative features from the audio signal, followed by the utilization of these features as input to a binary classifier for discerning between genuine and artificially generated audio. The end-to-end approach optimizes the feature extractor and classifier jointly and integrate them into one model. 

Feature extraction in this context can be categorized into three main approaches. The first category involves using the short-time Fourier transform (STFT) on speech signals to derive short-term spectral features, including LPS~\cite{zhang2021lps} and LFCC~\cite{todisco2018lfcc}. The second category focuses on capturing long-range temporal information through methods such as those proposed by Tak \emph{et al.}~\cite{tak2020qcep} and Chettri \emph{et al.}~\cite{chettri2019ensemble}, utilizing constant-Q transform (CQT) and wavelet transform (WT) based features, respectively. In the third category, self-supervised learning techniques are employed to extract deep speech features, including wav2vec~\cite{schneider2019wav2vec}, XLS-R~\cite{babu2021xls}, and HuBERT~\cite{hsu2021hubert}, aiming to mitigate the challenge of obtaining costly training speech data. Recent developments have seen end-to-end approaches gaining prominence due to their competitive performance and avoidance of knowledge-based features, tailoring models for specific applications rather than generic decompositions. Prominent works in this domain include RawNet2~\cite{tak2021end}, leveraging sine cardinal filters, AASIST~\cite{jung2021AASISTAA}, which incorporates graph attention layers, and RawFormer~\cite{liu2023rawformer}, comprising both convolutional layers and transformer architectures.

While substantial advancements have been achieved in voice detection on in-domain datasets, there is a notable decrease in performance when applied to out-of-domain datasets~\cite{chen2020generalization,muller2022does}. This observation underscores a substantial challenge and emphasizes the necessity to enhance the generalizability of voice detectors. 
To enhance generalization, Xie \emph{et al.}~\cite{xie2023domain} advocate for aggregating genuine speech and discerning it from counterfeit audio by learning within a dedicated feature space. In a similar vein, Zhang \emph{et al.}~\cite{zhang2023you} employ a continual learning strategy that adaptively computes weight modifications based on the ratio of genuine to fake utterances, effectively mitigating catastrophic forgetting. The assessment of voice detector reliability is addressed by Salvi \emph{et al.}~\cite{salvi2023reliability}, demonstrating that focusing solely on the most pertinent segments of a signal can augment generalization capabilities. Wang \emph{et al.}~\cite{wang2023cross} propose a method to merge multi-view features for fake audio detection, selecting generalized features from prosodic, pronunciation, and wav2vec dimensions. In a different approach, Ding \emph{et al.}~\cite{ding2023samo} introduce a scheme that clusters authentic speech around multiple speaker attractors and repels spoofing attacks from all these attractors within a high-dimensional embedding space.

\section{Architecture}
\label{apx:architecture}
\smallskip
\noindent
\textbf{Encoder.} In our framework, there are two encoders with the same structure but different parameters: one generates content features, while the other produces domain features. Both are based on the RawNet2~\cite{tak2021end} structure with modifications, including the removal of RawNet2's last fully connected layer. The features before the last fully connected layer are used as the encoder outputs.

\smallskip
\noindent
\textbf{Header.} We employ a single-layer MLP (Multi-Layer Perceptron) as the classification output for both domain-agnostic and domain-specific tasks.

\smallskip
\noindent
\textbf{Decoder.} We concatenate domain-agnostic features and domain-specific features, then input them along with content features into the AdaIN (Adaptive Instance Normalization)~\cite{huang2017arbitrary} module for feature fusion. Finally, the combined features go through a series of convolutional layers and up-sample layers to obtain the audio output, illustrated as Table \ref{tab:appd:decoder-arch}
\begin{table}
\centering
\caption{The Decoder Architecture}
\vspace{0.1in}
\scalebox{1.0}{
	\begin{tabular}{cc}
	\hline
	\multicolumn{2}{c}{Input}                      \\ \hline
	\multicolumn{1}{c|}{$F_d$}        & $F_c$        \\ \hline
	\multicolumn{1}{c|}{reshape}     & reshape     \\ \hline
	\multicolumn{2}{c}{AdaIN}                      \\ \hline
	\multicolumn{2}{c}{upsample}                   \\ \hline
	\multicolumn{2}{c}{Conv2d + (Conv2d+Relu) x 2} \\ \hline
	\multicolumn{2}{c}{Conv2d +Upsample}           \\ \hline
	\multicolumn{2}{c}{Tanh}                       \\ \hline
	\end{tabular}

}
\label{tab:appd:decoder-arch}
\end{table}

\section{Loss implementation}\label{apx:mi_code}
To improve the universal applicability of domain-agnostic features across all voices, we maximize their mutual information by considering both joint and marginal distributions. Algorithm \ref{alg:code} is the key implementation of the Mutual Information Estimator, designed to compute the lower bound of mutual information.

\section{End-to-end Training Algorithm}\label{apx:algorithm_details}

Below is the pseudocode of our optimization with flattening loss landscape based on sharpness-aware minimization~\cite{foret2020sharpness}, and is implemented throughout the end-to-end training process.

\begin{algorithm}[tb]
\caption{Optimization with Flattening Loss Landscape}
\label{alg:algorithm}
\textbf{Input}: A training dataset $\mathcal{S}$ with domain variable $D$, num\_batch, max\_iterations, learning rate $\beta$, neighborhood size $\epsilon^*$.\\
\textbf{Output}: An AI-synthesized voice detector with generalizability.\\
\textbf{Initialization:} $\theta_0$, $l=0$
\begin{algorithmic}[1]
    \FOR{$e=1$ to \emph{max\_iterations}}
        \FOR{$b=1$ to \emph{num\_batch}}
            \STATE Sample a mini-batch $\mathcal{S}_b$ from $\mathcal{S}$
            \STATE Compute $\nabla_\theta \mathcal{L}$ and $\epsilon^*$ based on Eq. (\ref{eq:eq_sam})
            \STATE Compute gradient approximation for Eq. (\ref{eq:sharpness})
            \STATE Update $\theta$: $\theta_{l+1} \leftarrow \theta_{l} - \beta \nabla_\theta \mathcal{L}\big|_{\theta_l + \epsilon^*}$
            \STATE $l \leftarrow l + 1$
        \ENDFOR
    \ENDFOR
    \RETURN $\theta_{l}$
\end{algorithmic}
\end{algorithm}
\section{Datasets Description}\label{apx:dataset_description}

\subsection{Data information of training and testing set}\label{tab:data-info}

To assess the generalization of our method, we tested it on various mainstream audio benchmarks, including LibriSeVoc~\cite{sun2023ai}, WaveFake~\cite{frank2021wavefake}, ASVspoof 2019~\cite{lavrentyeva2019stc}, and the audio segment of FakeAVCeleb~\cite{khalid2021fakeavceleb}.
\textbf{1)} \textbf{\textit{LibriSeVoc}}\cite{sun2023ai} is a audio deepfake benchmark with 34.92 hours of real audio and 208.74 hours of synthesized voices generated using six widely-used vocoder structures: WaveNet~\cite{oord2016wavenet}, WaveRNN~\cite{kalchbrenner2018efficient}, WaveGrad~\cite{chen2020wavegrad}, DiffWave~\cite{kong2020diffwave}, MelGAN~\cite{kumar2019melgan}, and Parallel WaveGAN~\cite{yamamoto2020parallel}. 
\textbf{2)} \textbf{\textit{WaveFake}}~\cite{frank2021wavefake} is a audio deepfake detection benchmark with 196 hours of synthetic audio content generated using six vocoder architectures: MelGAN~\cite{kumar2019melgan}, FullBand-MelGAN~\cite{frank2021wavefake}, MultiBand-MelGAN~\cite{frank2021wavefake}, HiFi-GAN~\cite{kong2020hifi}, Parallel WaveGAN~\cite{frank2021wavefake}, and WaveGlow~\cite{prenger2019waveglow}.
\textbf{3)} \textbf{\textit{ASVspoof 2019}}~\cite{lavrentyeva2019stc} is built from the VCTK base corpus~\cite{veaux2016superseded}, integrating speech data from 107 speakers with 10 vocoders. 
\textbf{4)} \textbf{\textit{FakeAVCeleb}}~\cite{khalid2021fakeavceleb} is generated videos, whose voice segments are created by SV2TTS~\cite{jia2018transfer}.

\begin{table*}
\centering
\caption{Details of datasets.}
\vspace{0.1in}
\scalebox{1.0}{
    \begin{tabular}{c|c|c|c|c|c}
\hline
\textbf{Dataset} & \textbf{\# Vocoder type} & \textbf{Frequency} & \textbf{Training size} & \textbf{Dev size} & \textbf{Testing size} \\ \hline
LibriSeVoc & 6 & 24kHz & 55,440 & 18,480 & 18,487 \\
WaveFake & 6 & 16kHz & 64,000 & 16,000 & 24,800 \\
ASVspoof2019 & 10 & 16kHz & 25380 & 24844 & 71237 \\
FakeAVCeleb & 1 & 16kHz & - & - & 21544 \\ \hline
\end{tabular}
}

\label{tab:dataset-info}
\end{table*}

\subsection{Differences of mel-spectrogram between human voices and AI-synthesized}\label{apx:diff-mel-spec}
\begin{figure}
    \centering\includegraphics[scale=0.4]{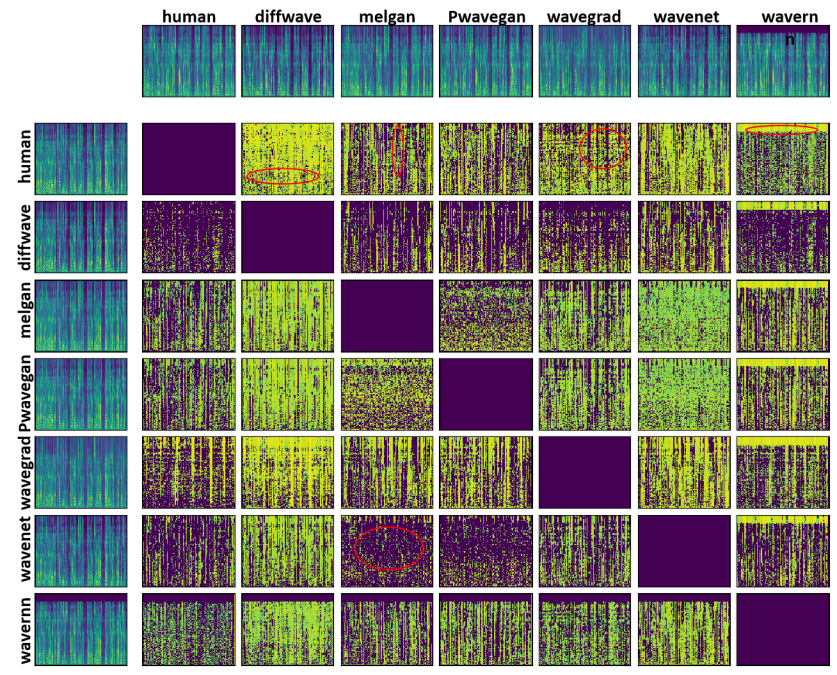}
    \caption{The artifacts introduced by the AI vocoders to a human voice signal. We show the mel-spectrogram of these signals and their mutual differences by making the left mel-spectrogram minus the right ones, respectively}
    \label{fig:motivation}
\end{figure}

We have observed that current detectors often concentrate on specific forgery artifacts. To illustrate this, we performed a statistical analysis on the mel-spectrograms of human voices and AI-synthesized voices, calculating their mutual differences as shown in Figure \ref{fig:motivation}. From this figure, it becomes evident that there are numerous forgery patterns easily detectable by detectors trained solely on these vocoders, resulting in inferior performance.

\begin{algorithm}[t]
\caption{MI (dv): PyTorch-like Pseudocode}
\label{alg:code}

\begin{lstlisting}[language=python]

def compute_mi_dv(c, a):
    # c: B x n0 x dim, features of `conetent`
    # a: B x n1 x dim, features of `domain-agnostic`
    B, n0, _ = c.shape
    B, n1, _ = a.shape 
    # B, n1, n0, B
    u = torch.mm(a, c.t())
    # B x B x n0 x n1
    u = u.reshape(B, n0, B, n1).permute(0, 2, 3, 1)
    mask = torch.eye(B).to(c.device)
    n_mask = 1 - mask
    
    # Compute the joint and margin score.
    E_joint = u.mean(2).mean(2)
    E_margin = log_sum_exp(u, 0) - math.log(u.size(0)).mean(2).mean(2)
    
    E_joint = (E_joint * mask).sum() / mask.sum()
    E_margin = (E_margin * n_mask).sum() / n_mask.sum()

    mi = E_joint - E_margin
    return mi

\end{lstlisting}
\end{algorithm}

\begin{table}[!h]
\caption{Overview of vocoders in datasets, `LSV', `ASP', `WF', and `FAVC' stand for LibriSeVoc, ASVspoof2019, WaveFake, and FakeAVCeleb, respectively.}
    \tablestyle{5pt}{1.1}
    \scalebox{1.0}{
        \begin{tabular}{c|cccc}
    \hline
    Vocoder & LSV & WF & ASP & FAVC \\ \hline
    DiffWave & \checkmark &  &  &  \\
    MelGAN & \checkmark & \checkmark &  &  \\
    Parallel WaveGAN & \checkmark & \checkmark &  &  \\
    WaveGrad & \checkmark &  &  &  \\
    WaveNet & \checkmark &  & \checkmark &  \\
    WaveRNN & \checkmark &  & \checkmark &  \\
    WaveGlow &  & \checkmark &  &  \\
    MultiBand-MelGAN &  & \checkmark &  &  \\
    FullBand-MelGAN &  & \checkmark &  &  \\
    HiFi-GAN &  & \checkmark &  &  \\
    Spectral &  &  & \checkmark &  \\
    Waveform &  &  & \checkmark &  \\
    WORLD &  &  & \checkmark &  \\
    Griffin-Lim &  &  & \checkmark &  \\
    MFCC\_vocoder &  &  & \checkmark &  \\
    Neural\_source\_filter &  &  & \checkmark &  \\
    STRAIGHT &  &  & \checkmark &  \\
    Vocaine &  &  & \checkmark &  \\
    SV2TTS &  &  &  & \checkmark \\ \hline
    \end{tabular}
    }
\label{tab:dataset-in-domainview}
\end{table}

\section{Additional Experimental Settings}\label{apx:additional_settings}
Our model is initialized using Kaiming initializers~\cite{he2015delving}. For optimization, we utilize the Adam optimizer~\cite{kingma2014adam} with a learning rate of 0.0002, a batch size of 16, and train for 50 epochs, selecting only the best checkpoint. Hyperparameters $\lambda_1$, $\lambda_2$, $\lambda_3$, and $\lambda_4$ are set to 0.1, 0.3, 0.05, and 0.03, respectively. The margin $b$ in $L_{con}$ is set to 3, and $\gamma$ in Eq. (\ref{eq:eq_sam}) is set to 0.07. Furthermore, the original voice signal is used as input, and the same data preprocessing as RawNet2~\cite{tak2021end} is applied, ensuring that all signals are padded to the same size (65536). For RawNet2, the learning rate is 0.0001, trained for 100 epochs with a weight decay of 1E-4, and a batch size of 32. For Sun \emph{et al.}~\cite{sun2023ai}, the weight decay is 0.0001, the batch size is 32, and training lasts for 50 epochs. Regarding XLS-R~\cite{babu2021xls} and WavLM~\cite{chen2022wavlm}, the learning rate is 0.0001, with a weight decay of 0.0001, a batch size of 16, and training extends to 50 epochs.

\subsection{Revisit Audio Deepfake Dataset in Domain-View}\label{tab:dataset-in-domain-view}
In our experiments, evaluating both intra-domain and cross-domain performance is pivotal. To achieve this, we revisit existing datasets, namely LibriSeVoc~\cite{khalid2021fakeavceleb}, WaveFake~\cite{frank2021wavefake}, ASVspoof2019~\cite{lavrentyeva2019stc}, and FakeAVCeleb~\cite{khalid2021fakeavceleb}. Subsequently, we partition these datasets into `seen' vocoder estimation sets and `unseen' vocoder estimation sets. The classification of `seen' and `unseen' is determined by the selected training set (LibriSeVoc or ASVspoof2019), as illustrated in Table \ref{tab:dataset-in-domainview}.





\end{document}